\RequirePackage{lineno}
\documentclass[superscriptaddress,amsmath,amssymb,nofootinbib]{revtex4}

\usepackage{graphicx}
\usepackage{dcolumn}
\usepackage{bm}
\usepackage{ulem}
\usepackage{xfrac}

\usepackage[printwatermark]{xwatermark}
\usepackage{xcolor}

\usepackage{graphicx,amsmath,amssymb}
\usepackage{epstopdf}
\usepackage{lineno}
\usepackage[colorlinks=true, urlcolor=blue, linkcolor=blue, citecolor=blue]{hyperref}

\begin{document}

\title{Effect of the field self-interaction of General Relativity on the Cosmic Microwave Background Anisotropies}

\author{ Alexandre Deur}

\affiliation{University of Virginia, Charlottesville, VA 22904. USA\\}
\affiliation{Old Dominion University, Norfolk, VA 23529. USA\\
\email{deurpam@jlab.org} 
}

\begin{abstract}

Field self-interactions are at the origin of the non-linearities inherent to General Relativity. 
We study their effects on the Cosmic Microwave Background anisotropies. We find that they
can reduce or alleviate the need for dark matter and dark energy in the description of
the Cosmic Microwave Background power spectrum.

\end{abstract}

\maketitle

\section{Introduction}

The power spectrum of the Cosmic Microwave Background (CMB) anisotropies is a leading
evidence for the existence of the dark components of the universe. This owes to the severely 
constraining precision of the observational data~\cite{CMB, Aghanim:2019ame} and to the
concordance within the dark energy-cold dark matter model ($\Lambda$-CDM, the standard model of 
cosmology) of the energy and matter densities obtained from the CMB with 
those derived from other observations, e.g.,  supernovae at large redshift $z$~\cite{large-z 98 data}. 
Despite the success of $\Lambda$-CDM, the absence of direct~\cite{Kahlhoefer:2017dnp} or 
indirect~\cite{Gaskins:2016cha} detection of dark matter particles is worrisome since
searches have nearly exhausted the parameter space where likely candidates could reside.
In addition, the straightforward extensions of particle physics' Standard Model, e.g.,  minimal SUSY, 
that provided promising dark matter candidates are essentially ruled out~\cite{Arcadi:2017kky}. 
$\Lambda$-CDM  also displays tensions with cosmological observations, e.g.,  it 
overestimates the number of dwarf galaxies and globular clusters~\cite{Klypin:1999uc}
or has no easy explanation for the tight correlations found between 
galactic dynamical quantities and the supposedly sub-dominant baryonic matter, e.g.,   
the Tully-Fisher~\cite{Tully:1977fu} or McGaugh {\it et al.} relations~\cite{McGaugh:2016leg}.
These worries are remotivating the exploration of alternatives to dark matter and possibly dark energy. 
To  be as compelling as $\Lambda$-CDM, such alternatives must explain 
the observations suggestive of dark matter/energy
consistently and economically ({\it viz}, without introducing many parameters and fields).
Among such observations, the CMB power spectrum is arguably the most prominent. 

Here we study whether the  self-interaction (SI) of gravitational fields, a defining 
property of General Relativity (GR), may
allow us to describe the CMB power spectrum without introducing dark components, 
or modifying the known laws of nature.
GR's SI already explains other key observations involving dark matter/energy: 
flat galactic rotation curves~\cite{Deur:2009ya},
large-$z$ supernova luminosities~\cite{Deur:2017aas},
large structure formation~\cite{Deur:2021ink}, and
internal dynamics of galaxy clusters, including the Bullet Cluster~\cite{Deur:2009ya}.
It also explains the Tully-Fisher and McGaugh {\it et al.} relations~\cite{Deur:2019kqi}.
First, we recall the origin of GR's SI and discuss when it becomes important 
as well as its overall effects. 

\section{Field self-interaction \label{sec:FSI}}
A crucial difference between Newtonian gravity and GR is that the former is a linear theory for which  
field superposition principle applies, while the latter is not, i.e., fields self-interact and the combination of two
fields differ from their sum.  
The origin of this phenomenon can be identified after analyzing the Lagrangian of GR,
\begin{eqnarray}
\mathcal{L}_{\mathrm{GR}}={\sqrt{\det(g_{\mu\nu})}\, g_{\mu\nu}R^{\mu\nu}}/{(16\pi G)}, 
\label{eq:Einstein-Hilbert Lagrangian}
\end{eqnarray}
where $G$  is the gravitational coupling,
$g_{\mu\nu}$ the metric, and
$R_{\mu\nu}$ the Ricci tensor. 
One defines the gravitational field $\phi_{\mu\nu}$ by the difference between $g_{\mu\nu}$ 
and a constant reference metric $\eta_{\mu\nu}$, e.g.,  that of Minkowski or Schwarzschild:
$\phi_{\mu\nu} \equiv (g_{\mu\nu} - \eta_{\mu\nu})/\sqrt{M} $. The normalization by 
 the system mass $M$ makes $\phi_{\mu\nu}$ the field due to a unit mass.
Developing $\mathcal{L}_{\mathrm{GR}}$ yields:
\vspace{-0.4cm}
\begin{eqnarray}
\mathcal{L}_{\mathrm{GR}}\!=\! \sum_{n=0}^\infty (16\pi MG)^{n/2}\left[\phi^n\partial\phi\partial\phi\right]. 
\label{eq:Polynomial Einstein-Hilber Lagrangian}
\end{eqnarray}
(We consider the pure field case, which is sufficient and simpler. 
The general case including matter can be found, e.g.,  in Refs.~\cite{Zee}.)
The $\left[\phi^{n}\partial\phi\partial\phi\right]$ denotes a sum of Lorentz-invariant terms of the form 
$\phi^{n}\partial\phi\partial\phi$.
Newtonian gravity is obtained by choosing the Minkowski metric for $\eta_{\mu\nu}$ and truncating 
Eq.~(\ref{eq:Polynomial Einstein-Hilber Lagrangian}) to $n=0$, 
with $\left[\partial\phi\partial\phi\right]=\partial^{\mu}\phi_{00}\partial_{\mu}\phi^{00}$ 
and $\partial^0\phi_{00}=0$. The $n>0$ terms, then, cause the field SI.
The same SI phenomenon exists with the nuclear Strong Force, which is formalized by 
quantum chromodynamics  (QCD). In fact, QCD and GR have the same lagrangian structure that, 
{\it inter alia}, enables fundamental field SI. Field SI is the hallmark of QCD since its large coupling
makes the SI effects prominent. 
In contrast in GR, field coupling is driven by $\sim \sqrt{GM/L}$ (with $L$ a length 
characterizing the system), whose typically small value allows us to   
approximate gravity with a linear theory, e.g.,  the Newtonian or the Fierz-Pauli approximations of GR. 
However, once $\sqrt{GM/L}$ becomes large enough, SI {\it must} arise. In fact, 
it was shown in Refs.~\cite{Deur:2009ya} that for galaxies, $\sqrt{GM/L}$ can be
sufficiently large to enable SI. 
These then strengthen the binding of the galaxy components in a manner that straightforwardly 
produces flat galactic rotation curves~\cite{Deur:2009ya}. The strengthening also alleviates the need for
dark matter to explain the growth of large structures~\cite{Deur:2021ink}. 
Employing Newtonian gravity to analyze these subjects 
overlooks the SI and, if the latter is important, induces apparent mass discrepancies 
interpreted as dark matter. 
Another crucial effect of SI arises from energy conservation: 
the strengthening of the system binding must be compensated by a suppression of the gravitational field outside of the system. 
For example in QCD, the increased binding confines quarks into nucleons and meanwhile the force is suppressed 
outside the nucleon resulting in a much weaker large-distance residual force, the Yukawa interaction.
If the equivalent effect~\cite{universe_Newton_evol} for massive systems bound by GR is overlooked, 
the large-distance suppression of gravity can be mistaken for a global repulsion (dark energy) 
that balances the supposedly pristine force~\cite{Deur:2017aas}. 
The direct connection between observations linked to dark energy and dark matter,
unexpected within $\Lambda$-CDM, naturally explains the cosmic coincidence problem~\cite{cosmic_coinc_pb}.

\begin{figure}
\center
\includegraphics[width=0.4\textwidth]{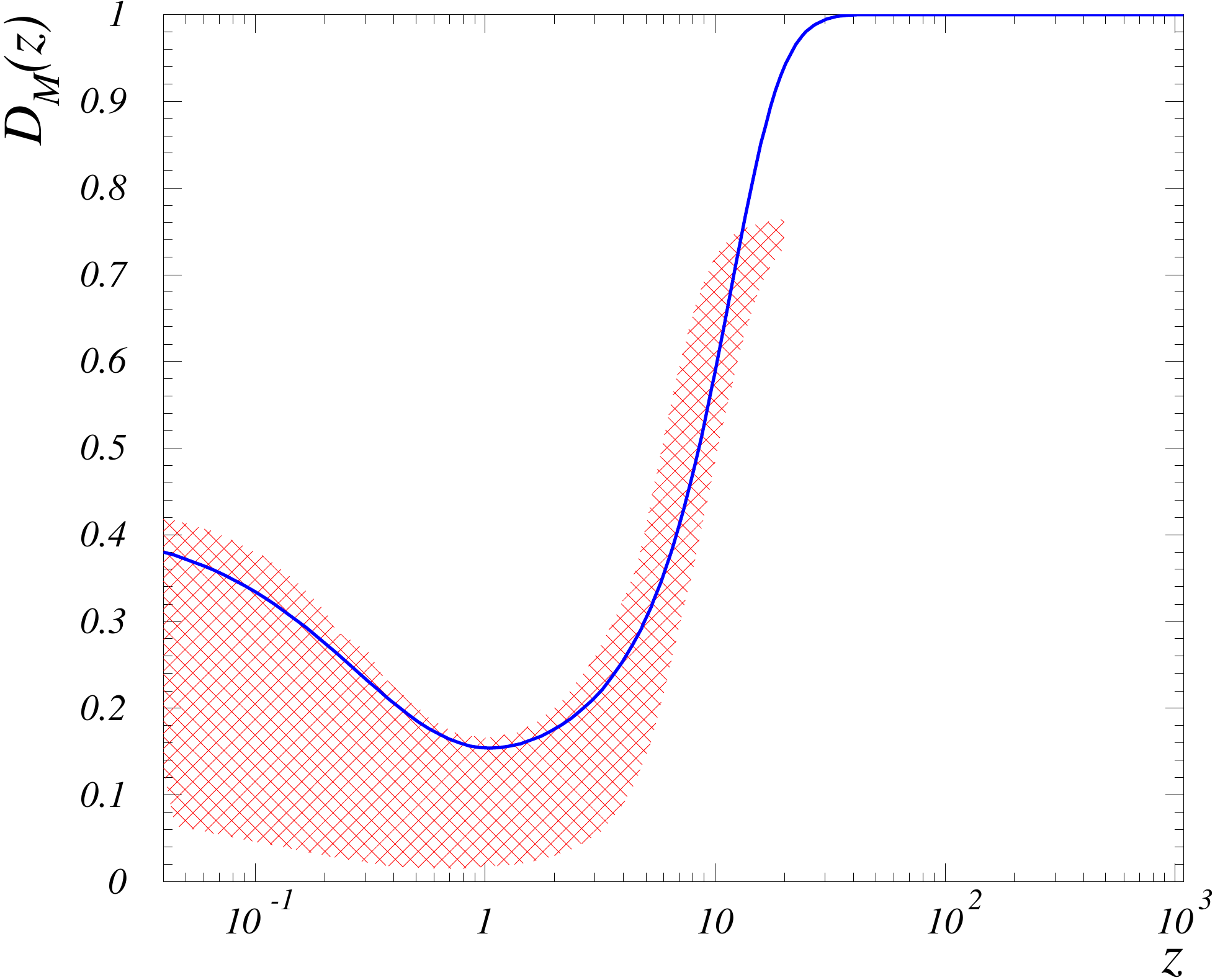}
\vspace{-0.4cm}
\caption{Depletion function $D_{M}$ vs redshift $z$. 
The line shows the $D_{M}(z)$ used in this work. It is constrained to remain within the $D_{M}(z)$ uncertainty 
determined in Ref.~\cite{Deur:2017aas} (hatched band).
}
\label{fig:D(z)}
\vspace{-0.5cm}
\end{figure}

The {\it local} effects of GR's SI, i.e., the increase of the binding energy of systems, 
can be directly calculated from Eq.~(\ref{eq:Einstein-Hilbert Lagrangian})~\cite{Deur:2009ya, Deur:2019kqi}.
{\it Global} effects, i.e.,  the large-distance suppression of gravity, have been treated effectively in Ref.~\cite{Deur:2017aas}
by folding them into a {\it depletion function} $D_M (z)$.
In fact,  lifting the standard Friedmann-Lema\^itre-Robertson-Walker (FLRW) 
approximations of  isotropy and homogeneity makes $D_M (z)$ 
to appear in the universe evolution equation~\cite{Deur:2017aas}.
There, $D_M=0$ represents  a full suppression of gravity at large-distance and $D_M=1$, none. 
In particular, $D_M(z) \approx 1$ at $z \gg 1$ since the early universe was 
nearly homogeneous and isotropic, while for the present web-like structured universe,
$D_M(z\approx0) < 1$. If all the fields were fully trapped into the structures that generate them, 
there would be no large-scale manifestation of gravity and $D_M(z\approx0) \to 0$. 
However, since the symmetry of a system suppresses SI effects ~\cite{Deur:2009ya, Deur:2013baa},
$D_M(z)$ increases at small $z$, e.g.,  because the ratio of elliptical over disk galaxies increases as 
$z\to0$~\cite{the:E/S ratio evolution3}. 
These features are conveniently parameterized by~\cite{Deur:2017aas}:
\vspace{-0.1cm}
\begin{equation}
D_M(z)= 1-(1+e^{(z-z_g)/\tau})^{-1}+Ae^{-z/b}
\label{eq:Dep_nom}
\end{equation} 
where $z_g$ is the redshift halfway through the galaxy formation epoch, 
$\tau$ the length of this epoch, 
$A$ the ratio of structures whose shapes evolved into more symmetric ones, and
$b$ the duration of this process. 
Fig.~\ref{fig:D(z)} shows the $D_M(z)$ from~\cite{Deur:2017aas}. 

Accounting for the consequences of GR's SI in the evolution of an inhomogeneous universe and of its large structures 
has been investigated using other methods than Eq.~(\ref{eq:Dep_nom}). Particularly,
backreaction effects, which also originate from field SI, have been proposed to explain without dark energy
the universe apparent  acceleration, see review~\cite{Schander:2021pgt}. 
Most such studies are performed perturbatively and therefore omit the nonperturbative effects that are 
crucial in analogous QCD phenomena, e.g., quark confinement or the feebleness of hadron-hadron interaction
comparatively to QCD's strength. From this perspective, it is unsurprising that these studies do not find important effects 
from backreaction. While perturbative treatments of backreaction are already nontrivial, nonperturbative ones are more difficult but indicate
that backreaction is important~\cite{Buchert:2015iva}.
Although backreaction studies and our approach both investigate the consequence for dark energy of the $n>0$ terms of 
Eq.~(\ref{eq:Polynomial Einstein-Hilber Lagrangian}), 
the former approach assumes dark matter whereas the latter deduces, by solving nonperturbatively Eq.~(\ref{eq:Polynomial Einstein-Hilber Lagrangian}), 
that dark matter is also a consequence of GR's SI~\cite{Deur:2009ya}. The latter approach then identifies an explicit nonperturbative process (field trapping) that 
exposes a direct connection between dark matter and dark energy~\cite{Deur:2017aas}. 
Aside investigations of backreation, other ideas have been proposed to explain the universe acceleration without dark energy, 
e.g., $f(R)$-gravity~\cite{DeFelice:2010aj}.
The key difference between these ideas and our approach is that they are beyond both GR, the current theory of gravity, 
and the Standard Model of Particle Physics (SMPP)
since they postulate new fields to be yet detected. For example, while Eq.~(\ref{eq:Polynomial Einstein-Hilber Lagrangian}) 
is only a reexpression of GR's Lagrangian, Eq.~(\ref{eq:Einstein-Hilbert Lagrangian}),~\cite{Zee}, 
$f(R)$ theories add to Eq.~(\ref{eq:Einstein-Hilbert Lagrangian}) terms in powers of $R \equiv g_{\mu\nu}R^{\mu\nu}$  that
go beyond GR, and requires ``scalaron'' fields and dark matter, both beyond the SMPP.
This contrasts with our approach that is within both  GR and the SMPP and connects dark energy and dark matter.  
Other extensions of GR than $f(R)$ theories may link dark energy to dark matter, e.g.,~\cite{Verlinde:2010hp} but so far they have not reproduced the  CMB. 

In the next section, we recall the expression of the CMB anisotropy correlation coefficient derived 
in the {\it hydrodynamic approximation}~\cite{Weinberg:2008zzc}. 
Then we discuss how SI modifies it and compare its computed values to the observations. In what follows,
$t$ denotes time,
$T$ the universe temperature, 
$a$ the Robertson-Walker scale factor, and 
$H$ the Hubble parameter, with $h$ its value in units of 100 km/s/Mpc.
The subscripts $0$, $EQ$ and $L$ 
indicate values for the present, matter-radiation equilibrium, and last scattering times, respectively.
$C^s_{TT,l}$ is the scalar multipole coefficient for the temperature-temperature angular correlation of the  CMB anisotropies,
with $l$ the multipole moment.
The baryon, total matter, radiation, photon, cold dark matter, and dark energy densities relative to the critical density
are $\Omega_B$, $\Omega_M$, $\Omega_R$, $\Omega_\gamma$, $\Omega_{DM}$, and $\Omega_\Lambda$, respectively,  
and $\Omega_K\equiv \sfrac{K}{a_0^2 H_0^2}$ with $K$ the metric curvature constant.
Since dark matter and dark energy are not assumed, $\Omega_{DM}= 0$ and $\Omega_\Lambda=0$ here.

\section{Effects of field self-interaction in the CMB anisotropies}
An analytical expression of $C^s_{TT,l}$ is convenient since it allows us to see 
where and how the SI  affects the CMB anisotropies.
We use the approximate expression derived in Ref.~\cite{Weinberg:2008zzc}:
\begin{eqnarray}
 \frac{l(l+1)C^s_{TT,l}}{2\pi}&=&\frac{4\pi T_0^2 N^2 e^{-2\tau_{reion}}}{25}\int_1^\infty d\beta \bigg(\frac{\beta l}{l_\mathcal{R}}\bigg)^{n_s-1} 
\bigg\{ \frac{3\sqrt{\beta^2-1}}{\beta^4 (1+R_L)^{\sfrac{3}{2}}} \mathcal{S}^2(\beta l /l_T) e^{-\sfrac{2\beta^2 l^2}{l_D^2}}
\sin^2\big(\beta l/l_H+\Delta(\beta l /l_T) \big) + \nonumber \\
&&  \frac{1}{\beta^2\sqrt{\beta^2-1}}\bigg[3\mathcal{T}(\beta l /l_T)R_L - (1+R_L)^{\sfrac{-1}{4}} \mathcal{S}(\beta l /l_T) e^{-\sfrac{\beta^2 l^2}{l_D^2}}
\cos\big(\beta l/l_H+\Delta(\beta l /l_T) \big)\bigg]^2 \bigg\}+\mathcal{C}(l),
\label{eq:hydro_approx}
\end{eqnarray}
where
$N$ is a factor normalizing the primordial perturbations, 
$\tau_{reion}$ is the optical depth of the reionized plasma,
$l_\mathcal{R}=(1+z_L)k_\mathcal{R}d_A$ with  $k_\mathcal{R}\equiv0.05$Mpc$^{-1}$ a conventional scale, and $d_A$ is the angular diameter distance of last scattering;
$n_s$ is the scalar spectral index, 
$R_L=\sfrac{3\Omega_B}{4\Omega_\gamma(1+z_L)}$,
$\mathcal{S},~\mathcal{T}$ and $\Delta$ are transfer functions,
$l_T=\sfrac{d_A}{d_T}$ with $d_T= \sfrac{\sqrt{\Omega_R}}{(1+z_L)H_0\Omega_M}$,
$l_D=\sfrac{d_A}{d_D}$ with $d_D$ the damping length,
$l_H=\sfrac{d_A}{d_H}$ with $d_H$ the acoustic horizon distance, and
$\mathcal{C}(l)$ is a second-order correction not present in Ref.~\cite{Weinberg:2008zzc}. 
The first term in the curly bracket includes the Doppler effect, while
the second term includes the Sachs-Wolf 
and intrinsic temperature anisotropy effects. Both terms include large-$l$ damping. 
Equation~(\ref{eq:hydro_approx}) is suited for the range  $30 \lesssim l \lesssim 2000$
since it does not include the integrated Sachs-Wolf, Sunyaev-Zel'dovich, and cosmic variance effects. 

Although Eq.~(\ref{eq:hydro_approx}) without $\mathcal{C}(l)$ provides a good overall description of the 
CMB anisotropies~\cite{Weinberg:2008zzc}, it is not fully accurate, hence 
the second-order correction $\mathcal{C}(l)$. It  is numerically obtained by the difference between the first-order term in Eq.~(\ref{eq:hydro_approx}) 
and a formally exact numerical calculation of $\sfrac{l(l+1)C^s_{TT,l}}{2\pi}$, e.g.~\cite{CAMB}, with both calculations 
performed with the $\Lambda$-CDM best fit parameters. 
It is sufficient for our purpose to restrict the SI corrections to the first-order 
term, since $\mathcal{C}(l)$ is by definition comparatively less dependent 
on cosmological parameters. 

At the time of last scattering, GR's SI effects are negligible. Thus, the mechanisms shaping  
Eq.~(\ref{eq:hydro_approx}) are unaffected and its form can be used as is. 
Since $\mathcal{S},~\mathcal{T}$, and $\Delta$ are time-independent and 
characterize the primordial scalar perturbations, the parameterizations~\cite{Weinberg:2008zzc} of these 
transfer functions also remain unmodified.
On the other hand, the cosmological parameters and characteristic scales (distance or multipole) entering Eq.~(\ref{eq:hydro_approx}) 
are derived from present-day $z=0$ values and evolved back to  $z=z_{L}$. It is through this evolution, which 
depends on the universe dynamics, that SI influences $C^s_{TT,l}$. The quantities affected are
the angular diameter distance of last scattering $d_A$ (Eq.~({\ref{Eq:dA})),
the damping length $d_D$, 
the scale $d_T$, 
the acoustic horizon distance  (Eq.~({\ref{Eq:dH})),
the $\Omega_i$, and
the ratio $R_L$.
Since $l_\mathcal{R}$, $l_T$,  $l_D$ and $l_H$ involve $d_A$, $d_T$, $d_D$ or $d_H$, 
they are also affected by SI.  

The distances $d_A$ and $d_H$ are given by:
 \vspace{-3mm}
 \begin{eqnarray}
d_A=\frac{1}{\sqrt{\Omega_K}H_0(1+z_L)}\sinh\bigg[\sqrt{\Omega_K} \int^1_{1/(1+z_L)} \frac{dx}{\sqrt{\Omega_\Lambda x^4+\Omega_K x^2 +\Omega_M x}}\bigg], \label{Eq:dA}
\end{eqnarray}
 \vspace{-3mm}
 \begin{eqnarray}
 d_H=\frac{2}{H_0(3R_L \Omega_M)^{\sfrac{1}{2}}(1+z_L)^{\sfrac{3}{2}}} \ln\big([\sqrt{1+R_L}+\sqrt{R_{EQ}+R_L}]/[1+\sqrt{R_{EQ}}]\big).
 \label{Eq:dH}
\end{eqnarray}

Two terms contribute to the damping length, $d_D \equiv \sqrt{d^2\textsubscript{Landau}+d^2 \textsubscript{Silk}}$:  
$$d^2 \textsubscript{Landau}=\frac{3\sigma^2 t_L^2}{8T_L^2 (1+R_L)},$$ 
where $\sigma$ is the standard deviation for the temperature $T_L$ owing to the fact  that recombination was not instantaneous, and:
 \vspace{-3mm}
$$d^2 \textsubscript{Silk}=\frac{R_L^2}{6(1-Y)(n_{B0})\sigma_\mathcal{T}H_0 \sqrt{\Omega_M}R_0^{\sfrac{9}{2}}} 
\int_0^{R_L} \frac{R^2 dR}{X(R)(1+R)\sqrt{R_{EQ}+R}}\bigg[\frac{16}{15}+\frac{R^2}{1+R}\bigg],$$ 
where $Y\simeq0.24$ is the density fraction for nucleons in neutral Helium, 
$n_{B0}$ the present baryon number density, $\sigma_\mathcal{T}$ the Thompson cross-section, $X$ the fractional ionization
of the plasma, and $R(t)\equiv \sfrac{3\rho_B(t)}{4\rho_\gamma(t)}$, 
with $\rho_i$ denoting average absolute densities. An approximation for $X$ is~\cite{jones:1985}:
\begin{eqnarray}
X(T)\simeq\bigg[ X(3400)^{-1}+\frac{\Omega_B}{\Omega_M^{1/2}} \int_T^{3400} \frac{84.2T'^{-0.1166}}{1+0.005085T'^{0.53}+42200T'^{0.8834}e^{-39474/T'}} dT' \bigg]^{-1}. \label{Eq:X(T)}
\end{eqnarray}
The SI affects $d\textsubscript{Silk}$ through
$\Omega_M$, $R_{EQ}$, and $X$ since the latter depends on $\Omega_M$ and $\Omega_B$.
The density ratio $R(t)$ is not affected by the SI because the SI leaves the
evolutions of $\rho_i$ unaffected~\cite{Deur:2017aas}. However, $R_{EQ}$ is affected 
since it is not defined as an absolute density ratio but as $R_{EQ}\equiv  \Omega_R R_0/\Omega_M$. 
$R_{EQ}$ is identical to $R(t_{EQ})$ for a FLRW universe, but not anymore once SI is accounted for. In this case it
must be redefined as:
$$R_{EQ}\equiv\frac{3\Omega_R\Omega_B}{4D_M(0)\Omega_\gamma(1+z_L)}.$$
The distance $d \textsubscript{Landau}$ is affected by SI only~\cite{note d_Landau} through the time of last rescattering, 
$$t_L=\frac{1}{H_0} \int_0^{1/(1+z_L)} x\big[\Omega_\Lambda x^4+\Omega_K x^2 +\Omega_M x +\Omega_R\big]^{-\sfrac{1}{2}}dx.$$ 

We now examine  how  SI specifically modifies the quantities just listed.
SI is accounted for by replacing 
the $\Omega_i$ by the {\it screened relative densities} 
$\Omega^*_M(z) \equiv \Omega_M D_M(z)$, $\Omega^*_B(z) \equiv \Omega_B D_M(z)$ 
and  $\Omega^*_K \equiv 1-\Omega_R-\Omega_M^*(z=0)$~\cite{Deur:2017aas}. 
 $\Omega_R$ and $\Omega_\gamma$ remain unscreened because radiation does not aggregate.
The $\Omega^*_i$ are 
effective dynamical quantities that enter in the evolution equation of the universe. Therefore, they
should not be compared with densities obtained from censuses or primordial synthesis. 

SI affects the parameters of Eq.~(\ref{eq:hydro_approx}) in two ways: \\
(1) it enhances local gravitational attraction;\\%
(2) it globally suppresses gravity at large distances.\\
Effect (1) is important when the local universe density variation is large, i.e., for $t \gg t_L$. 
Since the mechanisms producing Eq.~(\ref{eq:hydro_approx}) occurred when density variation was small
(the integrated Wolf-Sachs and Sunyaev-Zel'dovich effects are not included in Eq.~(\ref{eq:hydro_approx})), 
effect (1) can be ignored. In other words, in the expressions formalizing  the mechanisms generating the temperature anisotropies,
$\Omega_B$ can remain since $\Omega^*_B(z\gg1) \simeq \Omega_B$. 
Effect (2) influences the universe evolution and therefore affects how characteristic scales evolved since $t_L$. 
Thus, when $\Omega_M$ enters the expressions related to the evolution of the universe,  it is replaced by $\Omega^*_M$.
Since $\Omega_M$ is defined relative to the critical density for a FLRW universe,
for a flat universe with $\Omega_\Lambda=0$ and $\Omega_R\ll 1$,  
$\Omega_M=1$ and $\Omega^*_M=D_M(z)$. 
(Note that for effect (2), $\Omega_B$ is irrelevant since it is not explicitly present in the universe evolution equation, 
being included in $\Omega_M$). 
Table~\ref{tab:param_mod} summarizes the expressions modified to account for SI.
\begin{table}[t]
\caption{Expressions of the parameters explicitly affected by GR's SI. 
Column 1: denominations.
Column 2: standard expressions (FLRW universe).
Column 3: expressions accounting for SI.
For $X(T$), $g(T')$ is the integrand in Eq.~(\ref{Eq:X(T)}) and is unaffected by SI. 
The expressions for other parameters, such as characteristic multipoles or lengths,
are not modified although their values change since their expressions involve quantities listed in the table.
}
\label{tab:param_mod}
\resizebox{1.\textwidth}{!}{%
\begin{tabular}{|c|c|c|} 
\hline
   &   FLRW universe          &    Universe with GR's SI accounted for         \\  \hline \hline 
$d_A$  & $\frac{1}{\sqrt{\Omega_K}H_0(1+z_L)}\sinh\big[\sqrt{\Omega_K} \int^1_{1/(1+z_L)} \frac{dx}{\sqrt{\Omega_\Lambda x^4+\Omega_K x^2 +\Omega_M x}}\big]$  &
$\frac{1}{\sqrt{\Omega^*_K}H_0(1+z_L)}\sinh\big[\sqrt{\Omega^*_K} \int^1_{1/(1+z_L)} \frac{dx}{\sqrt{\Omega^*_K x^2 +D_M(1/x-1) x}}\big]$  \\ \hline 	
$d_T$   & $\sqrt{\Omega_R}/[(1+z_L)H_0\Omega_M]$  &     $\sqrt{\Omega_R}/[(1+z_L)H_0 D_M(0)]$        \\ \hline 	
$d_H$  &  $\frac{2}{H_0(3R_L \Omega_M)^{\sfrac{1}{2}}(1+z_L)^{\sfrac{3}{2}}} \ln([\sqrt{1\hspace{-1mm}+\hspace{-1mm}R_L}\hspace{-1mm}+\hspace{-1mm}\sqrt{R_{EQ}\hspace{-1mm}+\hspace{-1mm}R_L}]/[1\hspace{-1mm}+\hspace{-1mm}\sqrt{R_{EQ}}])$
  & $\frac{2}{H_0(3R_L D_M(0))^{\sfrac{1}{2}}(1+z_L)^{\sfrac{3}{2}}} \ln([\sqrt{1\hspace{-1mm}+\hspace{-1mm}R_L}\hspace{-1mm}+\hspace{-1mm}\sqrt{R_{EQ}\hspace{-1mm}+\hspace{-1mm}R_L}]/[1\hspace{-1mm}+\hspace{-1mm}\sqrt{R_{EQ}}])$       \\ \hline 	
$R_L$ &  $[3\Omega_B]/[4\Omega_\gamma(1+z_L)]$  &   $[3\Omega_R\Omega_B]/[4D_M(0)\Omega_\gamma(1+z_L)]$          \\ \hline 	
$t_L$  & $\frac{1}{H_0} \int_0^{1/(1+z_L)} x[\Omega_\Lambda x^4+\Omega_K x^2 +\Omega_M x +\Omega_R]^{-\sfrac{1}{2}}dx$  &
$\frac{1}{H_0} \int_0^{1/(1+z_L)} x[\Omega^*_K x^2 +D_M(1/x-1)+\Omega_R]^{-\sfrac{1}{2}}dx$  \\ \hline 	
$d^2 \textsubscript{Silk}$   & $\frac{R_L^2}{6(1-Y)(n_{B0})\sigma_\mathcal{T}H_0 \sqrt{\Omega_M}R_0^{\sfrac{9}{2}}} 
\int_0^{R_L} \frac{R^2 dR}{X(R)(1+R)\sqrt{R_{EQ}+R}}\big[\frac{16}{15}\hspace{-1mm}+\hspace{-1mm}\frac{R^2}{1+R}\big]$  &       
 $\frac{R_L^2}{6(1-Y)(n_{B0})\sigma_\mathcal{T}H_0 \sqrt{D_M(0)}R_0^{\sfrac{9}{2}}} 
\int_0^{R_L} \frac{R^2 dR}{X(R)(1+R)\sqrt{R_{EQ}+R}}\big[\frac{16}{15}\hspace{-1mm}+\hspace{-1mm}\frac{R^2}{1+R}\big]$   \\ \hline 	
$X(T)$  &   $1/\big[ X^{-1}(3400)+\frac{\Omega_B h^2}{(\Omega_M h^2)^{1/2}} \int_T^{3400} g(T')dT' \big]$ &       
$1/\big[ X^{-1}(3400)+\frac{\Omega_B h^2}{(D_M(0) h^2)^{1/2}} \int_T^{3400} g(T')dT' \big]$     \\ \hline 	
\end{tabular}}
\end{table}

We can now compute $C^s_{TT,l}$ with the SI effects included. 
The values of $T_0$, $\Omega_\gamma$ and $\Omega_R$ 
are well known from the measurement of the CMB average 
temperature. We use $T_0=2.7255$K 
\cite{PDG2020}, $\Omega_\gamma h^2=2.47\times 10^{-5}$ 
and $\Omega_R = 1.6813\Omega_\gamma $~\cite{Weinberg:2008zzc}. 
The remaining quantities ($H_0$, $n_s$, $z_L$ (or $T_L$), $\sigma_{t_L}$, $\Omega_B$ and $Ne^{-\tau_{reion}}$) 
are the free parameters to be determined by fitting the CMB data.  
The number of parameters is lower than for $\Lambda$-CDM since $\Omega_{DM}\equiv0$, $\Omega_\Lambda\equiv0$ and 
no assumption is needed for the thermodynamical properties of dark energy. 
In principle, $D_M (z)$, see Eq.~(\ref{eq:Dep_nom}), is  fixed but in practice it can vary within its uncertainty, see Fig.~\ref{fig:D(z)}. 
We used $z_g=8.5$, $\tau=4.5$, $A=0.25$ and $b=0.3$. 
Equation~(\ref{eq:hydro_approx}) with the expressions in the third column of Table~\ref{tab:param_mod} and 
$D_M(z)$ from Eq.~(\ref{eq:Dep_nom}) fits the scalar CMB anstropies for
$\Omega_B = 0.026$, 
$H_0=70$km/s/Mpc, 
$Ne^{-\tau_{reion}}=1.24\times10^{-5}$,  
$n_s=0.97$, 
$z_L=1380$ and 
$\sigma / T_L = 0.17$,
see Fig.~\ref{fig:Pwr_spectrum}. 
\begin{figure}
\center
\includegraphics[width=0.4\textwidth]{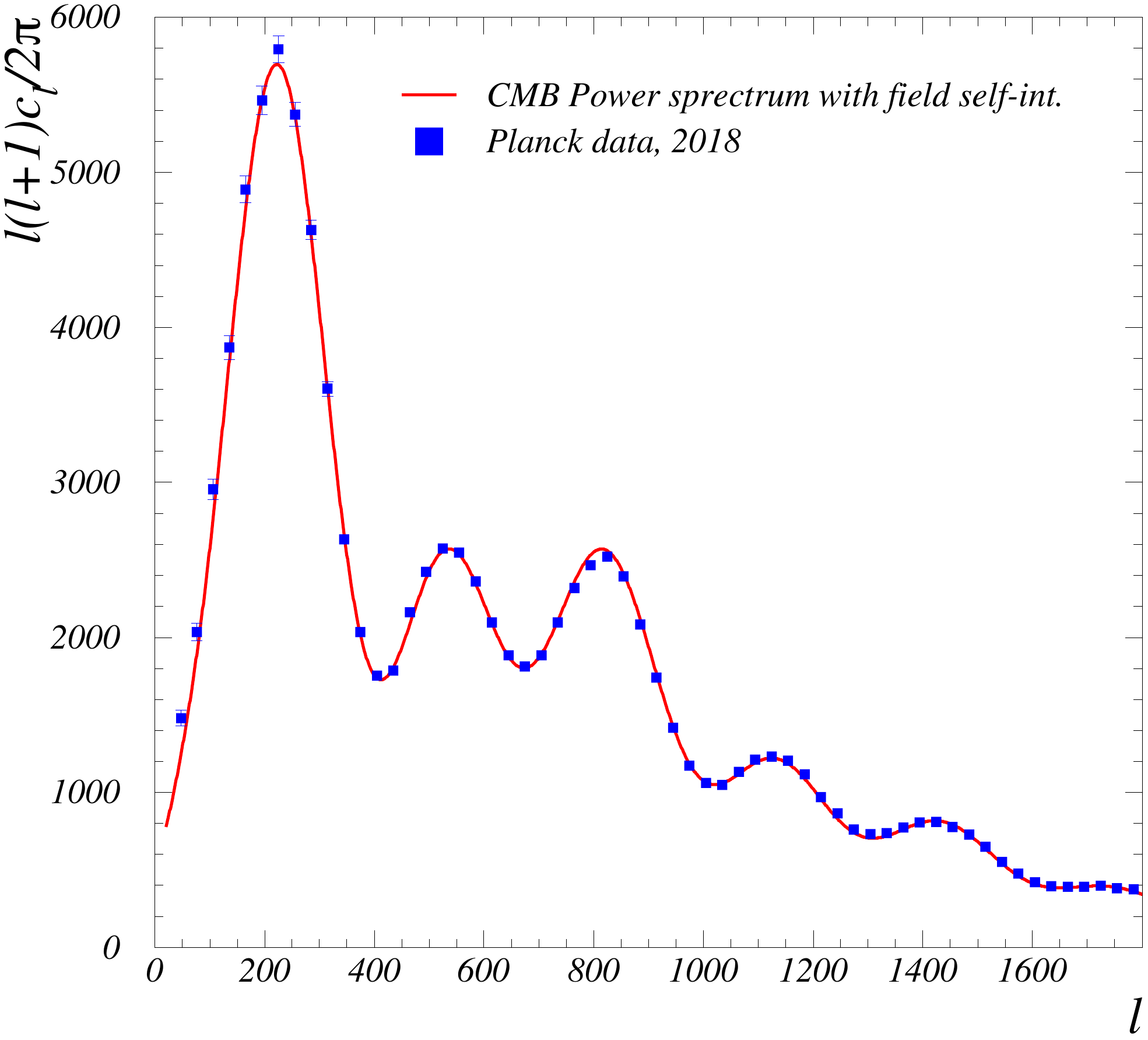}
\vspace{-0.4cm}
\caption{Power spectrum of the CMB  temperature anisotropy. The red line is the present calculation, 
to be compared to the measurement (squares, Planck 2018 release~\cite{Aghanim:2019ame}).}
\label{fig:Pwr_spectrum}
\vspace{-0.5cm}
\end{figure}
No uncertainties are assigned since our goal is not to
provide a precise determination of cosmological parameters, but rather to determine  if
SI can be a natural and viable alternative to $\Lambda$-CDM.

\section{Concordance with other observations}
To be plausible, an alternative to $\Lambda$-CDM must account consistently
for the observations related to dark matter and dark energy. Having examined the CMB anisotropies, the main object of this article, 
we now briefly discuss the consistency of the present approach with other important cosmological observations:
the apparent magnitude of standard candles at large-$z$, 
large structure formation,
the matter power spectrum $P(k)$,
and the age of the universe.
Except for $P(k)$, these observations were studied in detail in Refs.~\cite{Deur:2017aas,Deur:2021ink}
using the $D_M(z)$ originally determined in~\cite{Deur:2017aas} (red hatched band in Fig.~\ref{fig:D(z)}}).
Yet, although the band contains the $D_M(z)$ determined by our fit to the CMB observations (blue line  in Fig.~\ref{fig:D(z)}),
the latter is more constrained and therefore  need not provide results consistent with the other cosmological  observations
just mentionned.
Nevertheless, the large-$z$ data~\cite{large-z 98 data, SN data2, Schaefer}
remain well described with the 
$D_M(z)$ determined by our CMB fit, see Fig.~\ref{fig:spn_mps}, left panel.

The effect of SI on structure formation was treated in Ref.~\cite{Deur:2021ink}.
The central panel of Fig.~\ref{fig:spn_mps} shows the time-evolution of an 
overdensity  of initial value $\delta(t_L)=2 \times 10^{-5}$ calculated with the $D_M(z)$ from our fit in Fig.~\ref{fig:Pwr_spectrum}
and without dark matter. 
This yields $\delta(t_0) \approx 1$, consistent with the fact that structures had time to grow
to their present densities.

The observation of the present distribution of structures, expressed by $P(k)$ where $k$ is the wavenumber,
is another important constraint for cosmological models. 
As shown in the right panel of Fig.~\ref{fig:spn_mps}, the general shape of $P(k)$ 
can be described without dark matter/energy once SI effects are included. 
The reason  is that the effects of Inflation on the early universe are unaffected by SI since
the near uniformity and isotropy of the universe at early (post-inflation) times suppress SI
effects (see Section~\ref{sec:FSI}). Thus, at low $k$, $P(k) \propto k^{n_s}$ with 
$n_s=0.97$ from our CMB fit in Fig.~\ref{fig:Pwr_spectrum}.
This is valid up to $k \lesssim k_{max}$ with, in the SI framework,
$k_{max} = \sqrt{2}k_{EQ} = 2H_0 \sqrt{\Omega^*_M / a_{EQ}}$~\cite{Deur:2017aas}
(where the effect of $n_s \neq 1$ can be neglected:  $k_{max}$ increases by 1\% for $n_s=0.97$),
with $\Omega^*_M=D_M(0)$. 
Our CMB fit of Fig.~\ref{fig:Pwr_spectrum} 
yields  $k_{max}= 0.014$hMpc$^{-1}$, in agreement with observations~\cite{Chabanier:2019eai}. 
The spectrum maximum is given by 
$P(k_{max})=1.10\times 10^{14} [C(1/\Omega^*_M-1)N/h]^2/\Omega^*_M$
with $C(x)=\frac{5}{6}x^{-\sfrac{5}{6}}\sqrt{1+x}\int_0^x \frac{du}{u^{\sfrac{1}{6}}(1+u)^{\sfrac{3}{2}}}$~\cite{Weinberg:2008zzc}.
With our CMB fit values, $P(k_{max})=2.4 \times 10^4~(h^{-1}$Mpc$)^3$~\cite{note Pk_max}. 
The dissipative processes arising for $k > k_{max}$ involve in our case
only non-relativistic baryonic matter,
thus preserving the $P(k) \propto k^{-3}$ expected for $k>k_{max}$ 
from a nearly scale-invariant primordial power spectrum.
While the detailed analysis of the finer structure of  $P(k)$ is beyond 
the scope of this article, no difficulty is expected since that structure arises from 
baryon acoustic oscillations and Silk damping, both already well reproduced in Fig.~\ref{fig:Pwr_spectrum}.
\begin{figure}
\center
Large-$z$ standard candles
\hspace{1.7cm}
Large structure formation
\hspace{2.2cm}
Matter power spectrum~~~~ \par
\includegraphics[width=0.3\textwidth, height=0.3\textwidth]{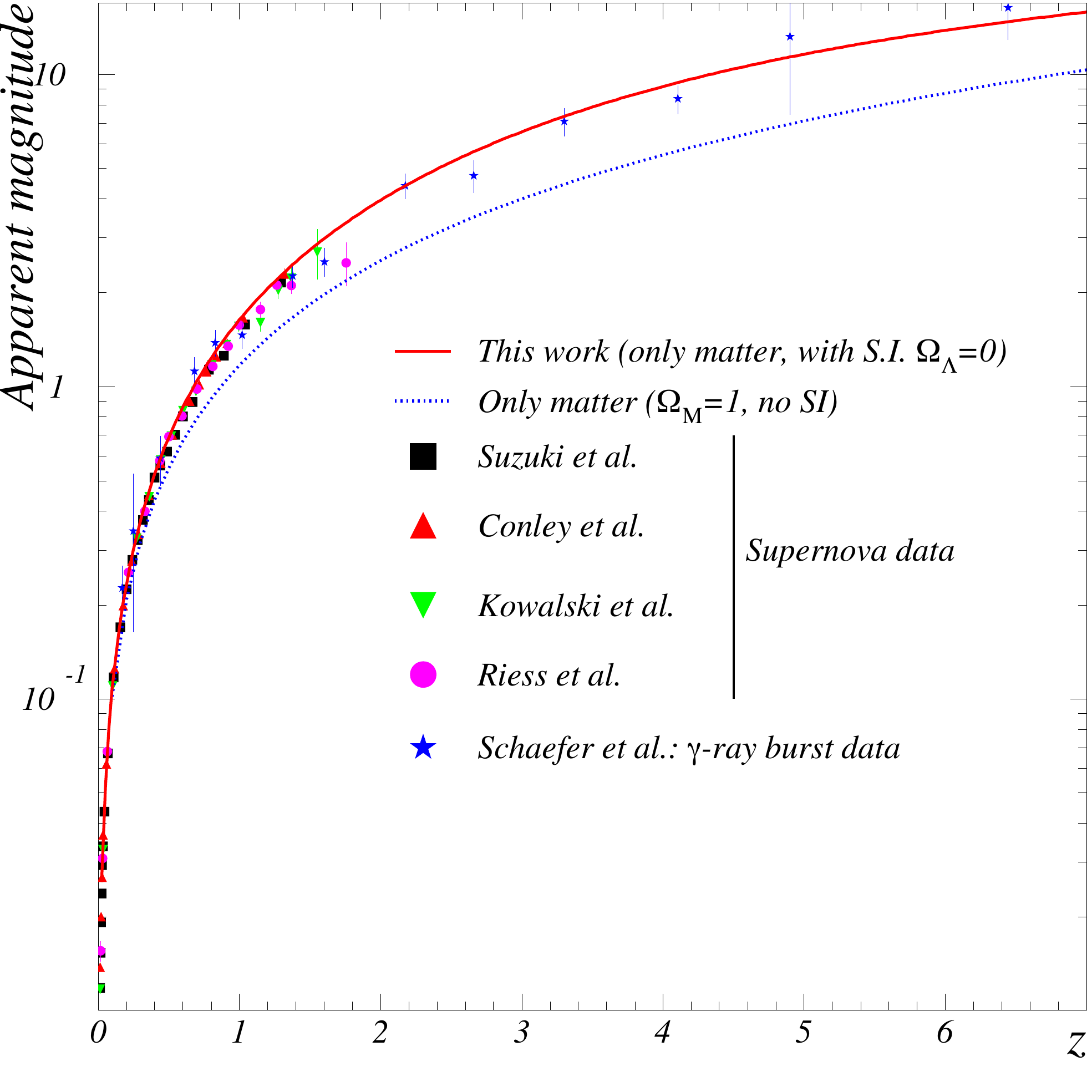}
\includegraphics[width=0.3\textwidth, height=0.3\textwidth]{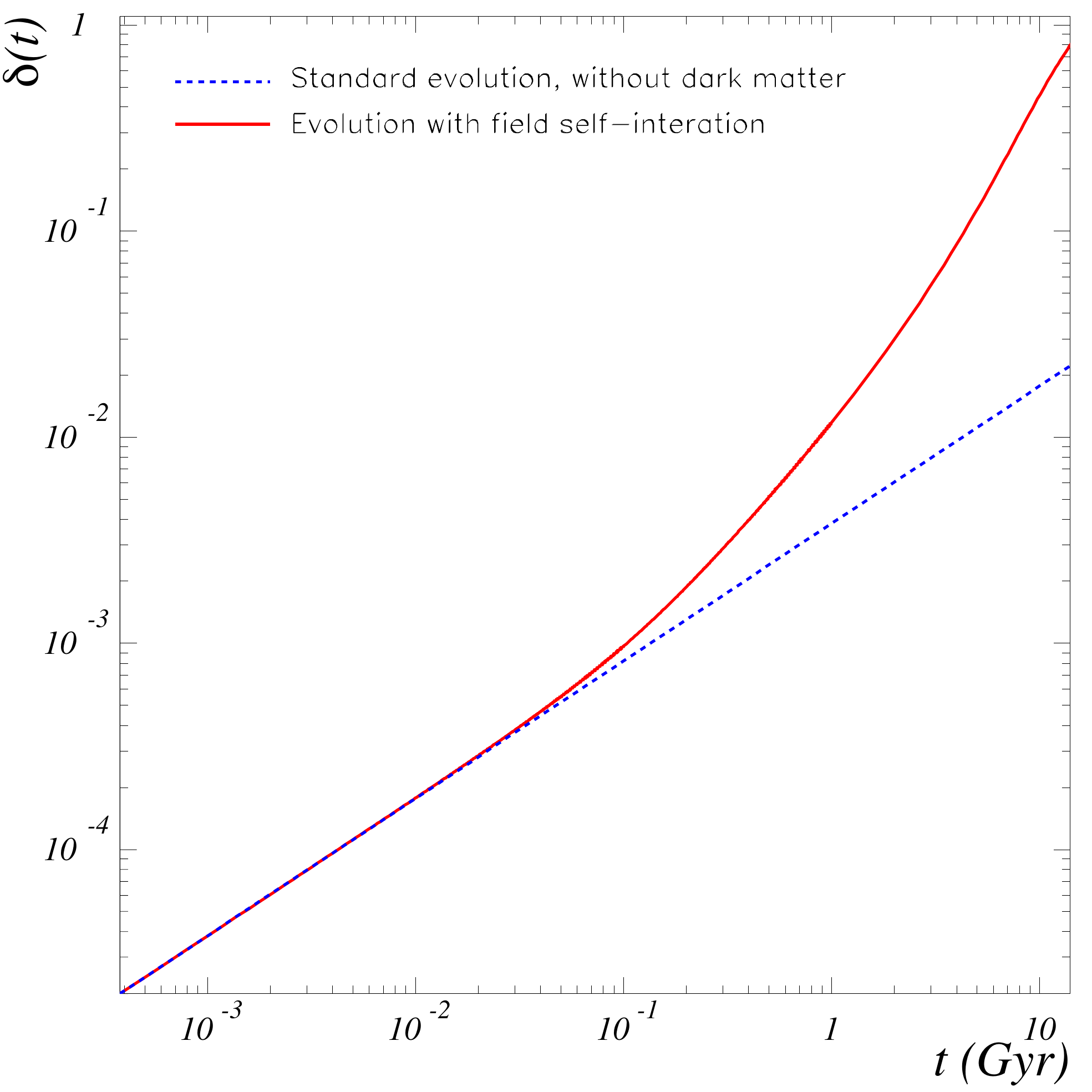}
\includegraphics[width=0.38\textwidth, height=0.3\textwidth]{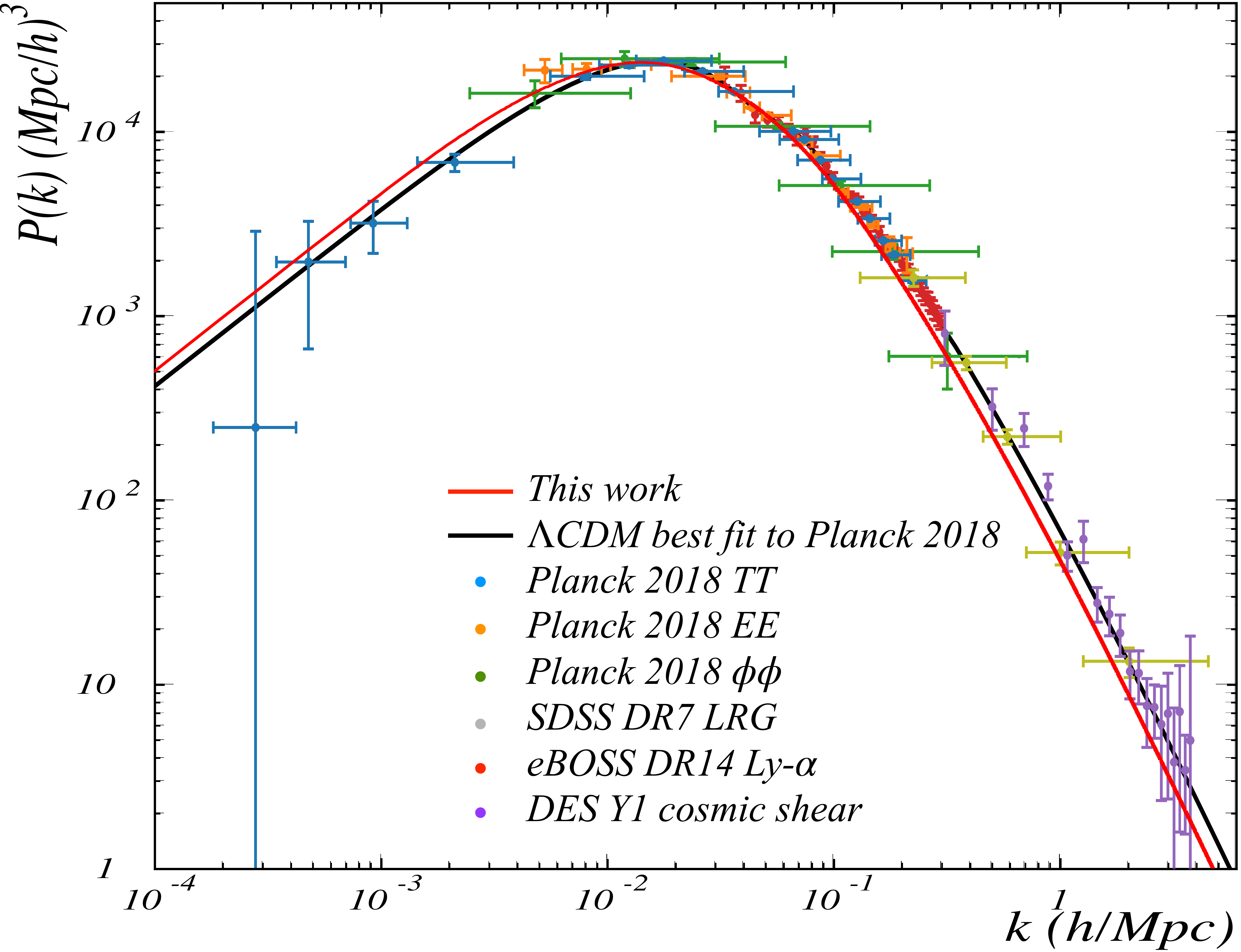}
\vspace{-0.3cm}
\caption{
Left panel: Supernova apparent magnitudes vs. redshift.  
Also shown are $\gamma$-ray burst data (star symbols).
The red line is the approach discussed here (universe containing only baryonic matter, with gravitational field self-interaction). 
The blue dotted line is for a flat FLRW universe with only matter. 
Central panel: Growth of an overdensity $\delta(t)$. Its initial value at the CMB
emission time, $t_L\approx 3\times 10^{-4}$ Gyr, is $2 \times 10^{-5}$. The red line shows the growth including
self-interaction effects. The dotted line is without them. 
Right panel: Matter power spectrum vs. wavenumber. The 
data points are from the 
Planck~\cite{Aghanim:2019ame}, 
SSDS (LRG~\cite{Reid:2010} and Li-$\alpha$~\cite{Chabanier:2018rga}),
and DES~\cite{DES:2017qwj}  data compiled in~\cite{Chabanier:2019eai}.
The red line is the approach discussed here using the first order calculation~\cite{Weinberg:2008zzc}. 
The black line is the $\Lambda$-CDM best fit to the Planck data. 
The three red curves are not fitted to the data shown in the panels: the curves' parameters are determined by
the function $D_M (z)$ obtained from the CMB fit of Fig.~\ref{fig:Pwr_spectrum}.
}
\label{fig:spn_mps}
\vspace{-0.5cm}
\end{figure}

Finally, the parameter values of our CMB fit in Fig.~\ref{fig:Pwr_spectrum} yield an age of the universe
after accounting for SI~\cite{Deur:2017aas} of 12.8Gyr. This agrees with the oldest 
known objects whose ages are estimated independently of cosmological models, namely the
chemical elements ($14.5^{+2.8}_{-2.5}$Gyr~\cite{Dauphas:2005}),
oldest stars ($13.5\pm1$Gyr~\cite{Catelan:2017}),
oldest star clusters ($12.8^{+0.9}_{-0.8}$Gyr~\cite{Kerber:2019}), and
oldest white dwarfs ($12.8\pm1.1$Gyr~\cite{Hansen:2004ih}).

\section{Conclusion}

Our result shows that the CMB scalar anisotropies may be accurately explained without dark 
matter and dark energy once the field self-interaction of General Relativity is accounted for. 
The effect of the latter is folded in a universal function $D_M (z)$ that is obtained from the 
timescale of large structure formation and the relative amounts  of baryonic matter present in
these structures~\cite{Deur:2017aas}.
$D_M (z)$ is universal since it explains high-$z$ supernovae measurements~\cite{Deur:2017aas} 
(Fig.~\ref{fig:spn_mps}, left panel), large structure formation~\cite{Deur:2021ink} (Fig.~\ref{fig:spn_mps}, central panel), 
the shape of the matter power spectrum (Fig.~\ref{fig:spn_mps}, right panel), and the CMB anisotropies (Fig.~\ref{fig:Pwr_spectrum}).
The SI of GR also eliminates the need for dark matter to explain the internal dynamics of structures: it 
straightforwardly yields flat rotation curves for disk galaxies~\cite{Deur:2009ya}, explains
the empirical tight relation between baryonic and observed accelerations~\cite{McGaugh:2016leg, Deur:2019kqi},
the Tully-Fisher relation~\cite{Tully:1977fu, Deur:2009ya, Deur:2017aas}, and
accounts for the dynamics of galaxy clusters~\cite{Deur:2009ya}.
This framework is economical --no exotic
matter, fields, nor modification of gravity are needed-- and natural. 
Interestingly, QCD, whose Lagrangian's structure is similar to that of GR, exhibits phenomena similar 
to those regarded as evidences for dark matter and dark energy. 

~\\
{ \bf Acknowledgments}
This work is funded in part by a Dominion Scholar grant from Old Dominion University.
The author thanks C. Sargent, S. \v{S}irca and B. Terzi\'c for useful discussions and for their comments on this article.


\begin{thebibliography}{00}
\bibitem{note d_Landau} Although $\sigma$ and $T_L$ are also affected by SI, 
the effect cancels in $d \textsubscript{Landau}$ because it only involves the ratio $\sigma/T_L$.
\end{thebibliography}
\end{document}